\documentclass[sigconf]{acmart}

\usepackage{longtable}
\usepackage[utf8]{inputenc} 
\usepackage[T1]{fontenc}    
\usepackage{hyperref}       
\usepackage{url}            
\usepackage{booktabs}       
\usepackage{amsfonts}       
\usepackage{nicefrac}       
\usepackage{microtype}      
\usepackage{xcolor}
\usepackage{colortbl}
\usepackage{etoc}
\usepackage{titletoc}
\usepackage{subfig}
\usepackage[most]{tcolorbox}
\usepackage{tikz} 
\usepackage{longtable}
\tcbuselibrary{breakable}
\usepackage{lipsum}
\usepackage{wrapfig}
\usepackage{dcolumn}
\usepackage{animate}
\usepackage{graphicx}
\usepackage{siunitx}
\usepackage[normalem]{ulem}
\usepackage{enumitem}

\usepackage{makecell}
\usepackage{multirow}
\usepackage{amsmath}
\usepackage{mathtools}
\usepackage{amsthm}
\usepackage{booktabs}
\usepackage{graphicx}
\usepackage{pifont} 
\newcommand{\cmark}{\textcolor{green!60!black}{\ding{51}}}  
\newcommand{\xmark}{\textcolor{red}{\ding{55}}}         

\usepackage{bbding}
\usepackage{fontawesome}
\usepackage{dcolumn}
\usepackage{makecell}
\usepackage{threeparttable}
\usepackage[font=small]{caption}

\AtBeginDocument{%
  }

\setcopyright{acmlicensed}
\copyrightyear{2018}
\acmYear{2018}
\acmDOI{XXXXXXX.XXXXXXX}
\acmConference[Conference acronym 'XX]{Make sure to enter the correct
  conference title from your rights confirmation email}{June 03--05,
  2018}{Woodstock, NY}
\acmISBN{978-1-4503-XXXX-X/2018/06}

\begin{document}

\title{FinMultiTime: A Four-Modal Bilingual Dataset for Financial Time-Series Analysis}

\author{Wenyan Xu}
\authornote{Equal contribution}
\affiliation{
  \institution{Central University of Finance and Economics}
  \country{}
}
\email{2022211032@email.cufe.edu.cn}

\author{Dawei Xiang}
\authornotemark[1]
\affiliation{
  \institution{University of Connecticut}
  \country{}
}
\email{ieb24002@uconn.edu}

\author{Yue Liu}
\affiliation{
  \institution{National University of Singapore}
  \country{}
}
\email{yliu@u.nus.edu}

\author{Xiyu Wang}
\affiliation{
  \institution{The University of Sydney}
  \country{}
}
\email{xiyuwang.usyd@gmail.com}

\author{Yanxiang Ma}
\affiliation{
  \institution{The University of Sydney}
  \country{}
}
\email{yama9404@uni.sydney.edu.au}

\author{Shu Hu}
\affiliation{
  \institution{Purdue University}
  \country{}
}
\email{hu968@purdue.edu}

\author{Liang Zhang}
\affiliation{
  \institution{The Hong Kong University of Science and Technology (Guangzhou)}
  \country{}
}
\email{liangzhang@hkust-gz.edu.cn}

\author{Chang Xu}
\affiliation{
  \institution{The University of Sydney}
  \country{}
}
\email{c.xu@sydney.edu.au}

\author{Jiaheng Zhang}
\affiliation{
  \institution{National University of Singapore}
  \country{}
}
\email{jhzhang@nus.edu.sg}


\begin{abstract}
 Pure time-series forecasting tasks typically focus exclusively on numerical features; however, real-world financial decision-making demands the comparison and analysis of heterogeneous sources of information. Recent advances in deep learning and large-scale language models (LLMs) have made significant strides in capturing sentiment and other qualitative signals, thereby enhancing the accuracy of financial time-series predictions. Despite these advances, most existing datasets consist solely of price series and news text, are confined to a single market, and remain limited in scale. In this paper, we introduce \textbf{FinMultiTime}, the first large-scale, cross-market multimodal financial time-series dataset. FinMultiTime temporally aligns four distinct modalities—financial news, structured financial tables, K-line technical charts, and stock price time series—across both the S\&P 500 and HS 300 universes. Covering 5,586 stocks from 2009 to 2025 in the United States and China, the dataset totals 112.6 GB and provides minute-level, daily, and quarterly resolutions, thus capturing short-, medium-, and long-term market signals with high fidelity. Our experiments demonstrate that (1) scale and data quality markedly boost prediction accuracy; (2) multimodal fusion yields moderate gains in Transformer models; and (3) a fully reproducible pipeline enables seamless dataset updates. The data for this paper can be found at\footnote{\url{https://huggingface.co/datasets/Wenyan0110/Multimodal-Dataset-Image_Text_Table_TimeSeries-for-Financial-Time-Series-Forecasting}}.
\end{abstract}




\maketitle

\section{Introduction}

Time-series regression models have long been the cornerstone of financial valuation and forecasting. Traditional statistical approaches \cite{weisberg2005applied, ariyo2014stock, bauwens2006multivariate, yang2025fourier} focus exclusively on numerical features and overlook open-domain knowledge from diverse modalities \cite{cheng2024sociodojo}.Intuitively, integrating multiple modalities enables richer, multidimensional representations that often outperform unimodal models \cite{huang2021makes}. In equity investment, for example, investors draw on historical price series and real-time multimodal data to inform buy, sell, or hold decisions: structured tables provide financial indicators, social media captures market sentiment, and technical charts reflect long-term price trends \cite{zhang2024multimodal, zou2023prebit}.

\begin{table*}[!t]
    \centering
    \caption{
    Comparison of existing multimodal financial time-series datasets. 
    }
    \vspace{-2mm}
    \renewcommand{\arraystretch}{1.3} 
    \resizebox{\linewidth}{!}{
    \begin{tabular}{lccccccccc}
        \toprule
        \textbf{Dataset Benchmarks} & \textbf{Venue \& Time} & \textbf{Domain} & \textbf{Language} & \textbf{Text} & \textbf{Time Series} & \textbf{Image} & \textbf{Table} & \textbf{Span} & \textbf{Time Interval} \\
        \midrule
        Time-MMD \citep{liu2024time} & NeurIPS 2024 & \multirow{2}{*}{\centering Multi-domain (Economics)} & English & \cmark & \cmark & \xmark & \xmark & 1989-2024 & Monthly \\
        CiK \citep{williams2024context} & arXiv 2024 & & English & \cmark & \cmark & \xmark & \xmark & 2024 & Monthly \\
        \cline{1-10}
        NewsForecast \citep{wang2024news} & NeurIPS 2024 & Multi-domain (Bitcoin) & English & \cmark & \cmark & \xmark & \xmark & 2019-2021 & Daily \\
        \cline{1-10}
        TimeCAP \citep{lee2025timecap} & AAAI 2025 & \multirow{2}{*}{\centering Multi-domain (Finance)} & English & \cmark & \cmark & \xmark & \xmark & 2019-2023 & Daily \\
        TSQA \citep{kong2025time} & arXiv 2025 & & English & \cmark & \cmark & \xmark & \xmark & -- & -- \\
        \cline{1-10}
        FNSPID Nasdaq \citep{dong2024fnspid} & KDD 2024 & \multirow{9}{*}{\centering Finance} & English, Russian & \cmark & \cmark & \xmark & \xmark & 2009-2023 & Minute-Level \\
        FTS-Text-MoE Nasdaq \citep{xu2025learning} & arXiv 2025 & & English & \cmark & \cmark & \xmark & \xmark & 2009-2025 & Minute-Level \\
        StockNet Dataset \citep{xu2018stock} & ACL 2018 & & English & \cmark & \cmark & \xmark & \xmark & 2014-2016 & Minute-Level \\
       CH-RNN Dataset \citep{wu2018hybrid} & CIKM 2018 & & English & \cmark & \cmark & \xmark & \xmark & 2017 & Minute-Level \\
        DOW30 \citep{chen2023chatgpt} & arXiv 2023 & & English & \cmark & \cmark & \xmark & \xmark & 2020-2022 & Daily \\
        TS-FF \citep{koval2024financial} & EMNLP 2024 & & English & \cmark & \cmark & \xmark & \cmark & 2010-2020 & Quarterly \\
        SEP \citep{koa2024learning} & WWW 2024 & & English & \cmark & \cmark & \xmark & \xmark & 2020-2022 & Minute-Level \\
        FinBen \citep{xie2024finben} & NeurIPS 2024 & & English, Spanish & \cmark & \cmark & \xmark & \xmark & -- & -- \\
        \textbf{FinMultiTime (Ours)} & -- & & English, Chinese & \cmark & \cmark & \cmark & \cmark & 2009-2025 & Minute-Level \\
        \bottomrule
    \end{tabular}
    }
    \vspace{-0.3cm}
    \label{table:dataset_comparison}
\end{table*}

\begin{figure*}[t!]
    \centering
    \begin{minipage}[b]{0.7\linewidth}
        \centering
        \vspace{3mm}
\includegraphics[width=\linewidth]{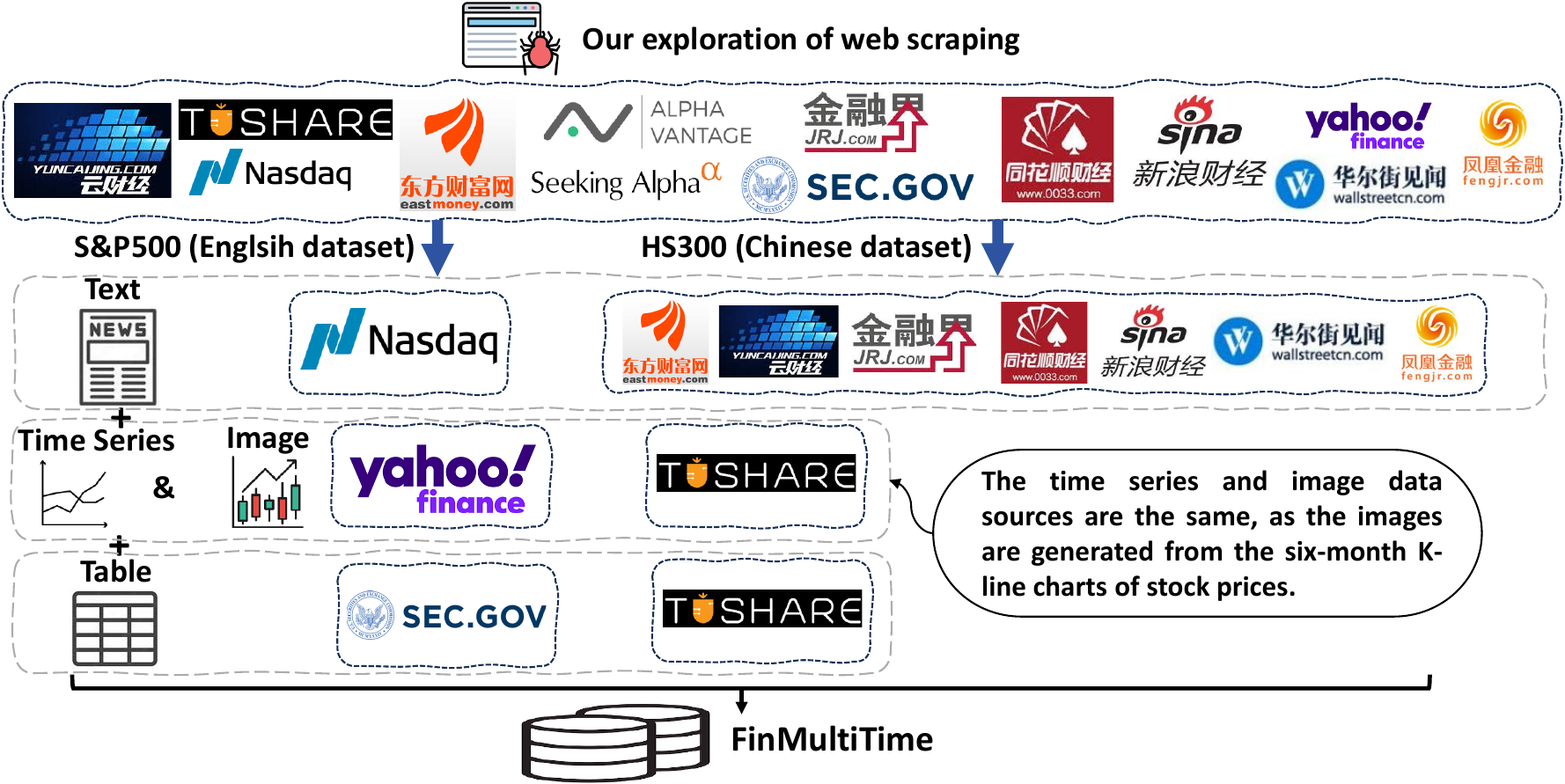}
        \caption{Data Collection Pipeline for the Bilingual Four-Modal FinMultiTime Dataset}
        \label{fig:Data Collection Pipeline}
    \end{minipage}
    \vspace{-3mm}
\end{figure*}

Moreover, according to the Efficient Market Hypothesis \cite{efficient_market}, prices absorb information with a lag, which provides a theoretical basis for exploiting multi-source signals not yet fully reflected in stock prices to predict future movements. Consequently, robust and reliable predictive models must assimilate heterogeneous data to capture the full complexity of price dynamics \cite{chai2020multi, fu2024incorporating}.

Recently, the natural language processing~(NLP) models enable sentiment analysis of financial news, event extraction from disclosures, table parsing in earnings reports, and automated chart summarization \cite{souma2019enhanced, araci2019finbert, yang2018dcfee, chapman2022towards, la2020end}.
Despite rapid advances in NLP models, existing multimodal datasets remain constrained. Most integrate only price and sentiment within a single market, risking information loss (Table \ref{table:dataset_comparison} ); recent efforts \cite{koval2024financial} incorporate quarterly tables but suffer from limited temporal coverage and low update frequency. Such datasets are too small to train large models or to validate generalization across market regimes \cite{liu2020finrl, mik2017smart, liu2023fingpt}, and they amplify the tendency of large language models (LLMs) to generate hallucinations in dynamic financial contexts \cite{gupta2023gpt, yang2023investlm}. 

To address these limitations, we introduce FinMultiTime, a large-scale bilingual dataset that spans from $2009$ to $2025$. FinMultiTime aligns temporal data across four modalities: text, tables, images, and time series. Our dataset includes $4694$ S\&P 500 constituents and $892$ HS 300 constituents. After rigorous cleaning and preprocessing, FinMultiTime comprises 112.6 GB of minute, daily and quarterly level data covering both U.S. and Chinese markets (Table \ref{Overview of Bilingual Financial Dataset}). Real-time updates ensure the dataset reflects the latest market conditions, providing a comprehensive foundation for developing and validating multimodal forecasting models. Experimental results demonstrate that incorporating large-scale multimodal data significantly reduces prediction error and improves trend-direction accuracy, with high-quality sentiment and long-term trend information proving especially critical. Our contributions lie in three aspects:
\begin{itemize}
  \item We first present \textbf{FinMultiTime}, a large-scale bilingual, cross-market, four-modality dataset for AI-driven stock market prediction (Table \ref{table:dataset_comparison} ).
  
  \item We provide reproducible and extensible usage guidelines to facilitate rapid adoption and expansion. The complete data collection, preprocessing pipeline, and example code will be available soon.
  
  \item Empirical results demonstrate that increasing dataset size significantly improves prediction performance (Table \ref{tab:sp500_hs300}), while enhancements in data quality and diversity further boost model accuracy (Table \ref{tab:hs300} and \ref{tab:sp500}).
\end{itemize}

\begin{table*}[ht]
\centering
\caption{Overview of Bilingual Financial Dataset Specifications for the HS300 (Chinese) and S\&P 500 (English) Indices}
\vspace{-3mm}
\resizebox{0.7\linewidth}{!}{%
\begin{tabular}{@{}lcccccc@{}}
\toprule
\textbf{Bilingual Dataset} & \textbf{Type} & \textbf{Size} & \textbf{Format} & \textbf{Stocks} & \textbf{Records} & \textbf{Frequency} \\ \midrule
\multirow{5}{*}{\textbf{HS300 (Chinese)}} 
 & Image        & 2.43 GB  & PNG         & 810   & 52,914    & Semi-Annual \\ 
 & Table        & 568 MB   & JSON/JSONL  & 810   & 2,430     & Quarterly/Annual \\ 
 & Time series  & 345 MB   & CSV         & 810   & 810       & Daily \\ 
 & Text         & 652.53 MB& JSONL       & 892   & 1,420,362 & Minute‑Level \\ 
 & All          & 3.96 GB       & --          & --    & 1,476,516        & -- \\ \midrule
\multirow{5}{*}{\textbf{SP500 (English)}} 
 & Image        & 8.67 GB  & PNG         & 4,213 & 195,347   & Semi-Annual \\ 
 & Table        & 84.04 GB & JSON/JSONL  & 2,676 & 8,028     & Quarterly/Annual \\ 
 & Time series  & 1.83 GB  & CSV         & 4,213 & 4,213     & Daily \\ 
 & Text         & 14.1 GB  & JSONL       & 4,694 & 3,351,852 & Minute‑Level \\ 
 & All          & 108.64 GB       & --          & --    & 3,559,440        & -- \\ \bottomrule
\end{tabular}}%
\label{Overview of Bilingual Financial Dataset}
\end{table*}

\section{Constructing FinMultiTime}
The construction of the FinMultiTime dataset begins with the systematic acquisition and processing of multi-source information. In this section, we detail the sources and procedures involved in assembling all modalities of FinMultiTime as shown in Figure~\ref{fig:Data Collection Pipeline}.

\begin{table*}[ht]
\centering
\caption{Comparison of Financial Tables for HS300 and S\&P 500. The 10-Q is a quarterly financial report filed by publicly traded companies, while the 10-K is a comprehensive annual report. Both provide detailed information on a company’s financial position, operating performance, and cash flow at the end of the reporting period.}
\vspace{-3mm} 
\renewcommand{\arraystretch}{1.2}
\resizebox{1\textwidth}{!}{
\begin{tabular}{lcccccc}
\hline
\multirow{2}{*}{\textbf{Table Type}} & \multicolumn{3}{c}{\textbf{HS300 (Chinese)}} & \multicolumn{3}{c}{\textbf{S\&P500 (English)}} \\
\cline{2-7}
 & Balance Sheet & Cash Flow Statement & Income Statement & Balance Sheet & Cash Flow Statement & Equity Statement \\
\hline
\textbf{Format} & JSONL & JSON & JSONL & JSON & JSONL & JSON \\
\hline
\textbf{Field Count} & 147 & 31 & 92 & 28 & 80 & 33 \\
\hline
\textbf{10-Q nums} & 48,537 & 81,070 & 45,257 & 81,070 & 47,526 & 81,070 \\
\hline
\textbf{10-K nums} & 24,551 & 27,793 & 18,260 & 27,793 & 18,636 & 27,793 \\
\hline
\textbf{Time span} & \multicolumn{3}{c}{\centering 2001/12/31-2024/09/30} & \multicolumn{3}{c}{\centering 2000/01/03-2025/04/25} \\
\hline
\label{table:Financial Tables Comparison}
\end{tabular}
}
\end{table*}

\begin{table*}[!t]
\centering
\caption{Comparison of Two News Sources and Data Attributes}
\renewcommand{\arraystretch}{1.2} 
\resizebox{0.98\linewidth}{!}{
\begin{tabular}{lcccccccc}
\hline
\textbf{Source} & \textbf{Nasdaq News} & \textbf{Sina Finance} & \textbf{WallstreetCN} & \textbf{10jqka} & \textbf{Eastmoney} & \textbf{Yuncaijing} & \textbf{Fenghuang} & \textbf{Jinrongjie} \\ \hline
\textbf{Time Period} & 2009-04-08 to 2025-04-08 & \multicolumn{7}{c}{2020-03-31 to 2025-03-31} \\ \hline
\textbf{Stock Symbol} & {\textcolor{blue!90}{Yes}} & {\textcolor{red!90}{No}} & {\textcolor{red!90}{No}} & {\textcolor{red!90}{No}} & {\textcolor{red!90}{No}} & {\textcolor{red!90}{No}} & {\textcolor{red!90}{No}} & {\textcolor{red!90}{No}} \\ \hline
\textbf{Headline} & {\textcolor{blue!90}{Yes}} & {\textcolor{red!90}{No}} & {\textcolor{blue!90}{Yes}} & {\textcolor{blue!90}{Yes}} & {\textcolor{blue!90}{Yes}} & {\textcolor{blue!90}{Yes}} & {\textcolor{red!90}{No}} & {\textcolor{blue!90}{Yes}} \\ \hline
\textbf{URL} & {\textcolor{blue!90}{Yes}} & {\textcolor{red!90}{No}} & {\textcolor{red!90}{No}} & {\textcolor{red!90}{No}} & {\textcolor{red!90}{No}} & {\textcolor{red!90}{No}} & {\textcolor{red!90}{No}} & {\textcolor{red!90}{No}} \\ \hline
\textbf{Text Type} & {Article} & \multicolumn{7}{c}{\centering {Flash News}} \\ \hline
\textbf{Filter Rate} & -- & 18.12\% & 14.83\% & 22.51\% & 21.20\% & 53.39\% & 19.57\% & 24.35\% \\ \hline
\textbf{Summarization} & {\textcolor{red!90}{No}} & {\textcolor{blue!90}{Yes}} & {\textcolor{blue!90}{Yes}} & {\textcolor{blue!90}{Yes}} & {\textcolor{blue!90}{Yes}} & {\textcolor{blue!90}{Yes}} & {\textcolor{blue!90}{Yes}} & {\textcolor{blue!90}{Yes}} \\ \hline
\textbf{Language} & English & Chinese & Chinese & Chinese & Chinese & Chinese & Chinese & Chinese \\ \hline
\end{tabular}
}
\label{Two News Sources}
\end{table*}

\subsection{Data Mining} 
\begin{figure*}[t!]
    \centering
    \begin{minipage}[b]{0.49\linewidth}
        \centering
        \includegraphics[width=\linewidth]{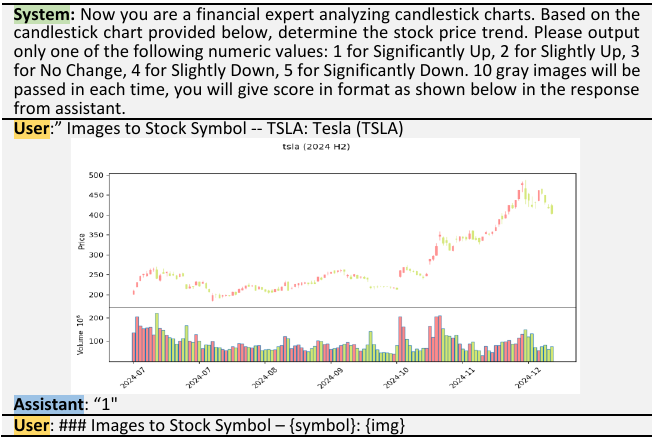}
        \vspace{-3mm} 
        \caption{Prompt–Response Example for Candlestick Chart Six-Month Trend Scoring}
        \label{fig:image_prompt}
    \end{minipage}
    \hfill
    \begin{minipage}[b]{0.49\linewidth}
        \centering
        \includegraphics[width=\linewidth]{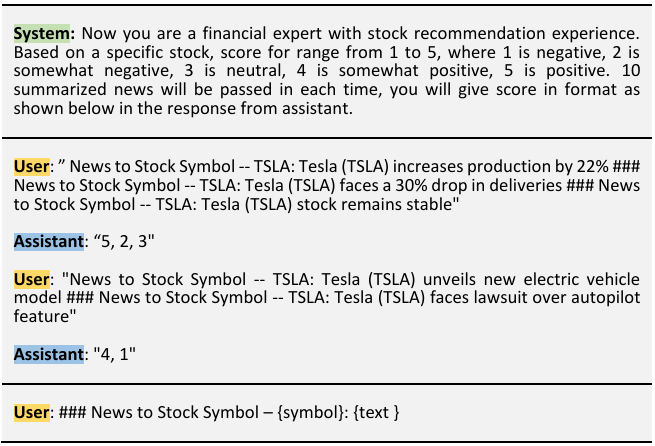}
        \vspace{-3mm} 
        \caption{Prompt–Response Example for Assigning 1–5 Sentiment Scores to News Items}
        \label{fig:news_prompt}
    \end{minipage}
    \vspace{-3mm}
\end{figure*}

We collect data from two of the major financial markets, as shown in Table \ref{Overview of Bilingual Financial Dataset}. For \textbf{the U.S}. stock \textbf{numerical} data, we first retrieve daily OHLCV (\textit{Open, High, Low, Close, and Volume}) data for S\&P 500 constituent stocks via the Yahoo Finance API \footnote{\scriptsize\url{https://finance.yahoo.com/}}. We segment the data into semi-annual windows and visualize it using candlestick \textbf{charts} generated with the \texttt{mplfinance} library. The upper panel plots daily prices, where red candlesticks indicate rising prices (\textit{close}~$>$~\textit{open}), and green candlesticks indicate falling prices. The lower panel shows daily trading volume, with bar heights representing volume in millions and colors aligned with price direction, illustrating the relationship between price movement and trading activity (see Figure \ref{fig:image_prompt} Tesla 2024 H2 chart). For the \textbf{news} sentiment data, we initially explored several platforms (e.g., Investing.com, Seeking Alpha, and Alpha Vantage), but strict usage restrictions limited their accessibility. Inspired by the FNSPID project \cite{dong2024fnspid}, we adopted a strategy of scraping publicly available news from Nasdaq. Building on the original FNSPID scripts, we developed a more robust, continuously running pipeline with several enhancements, including improved handling of abnormal pages, refined auto-pagination, cookie popup filtering, and compatibility with multiple ChromeDriver versions. The scraping process comprises two phases: the first uses Selenium to collect news headlines and corresponding URLs for each stock; the second retrieves the full article content from these URLs. The extracted texts constitute the news modality of our dataset. Structured financial \textbf{tables} are obtained primarily via the Securities and Exchange Commission (SEC) Submissions and Company Facts APIs \footnote{\scriptsize\url{https://www.sec.gov/search-filings/edgar-application-programming-interfaces}}. From 10-K and 10-Q filings of S\&P 500 companies since 2000, we automatically extract key indicators from XBRL facts in balance sheets, cash flow statements, and statements of shareholders’ equity, while removing irrelevant fields such as announcement dates and filing types.  For details on the U.S. financial data, see Table \ref{table:Financial Tables Comparison}.

\begin{figure*}[t!]
    \centering
    \begin{minipage}[b]{1\linewidth}
        \centering
        \includegraphics[width=\linewidth]{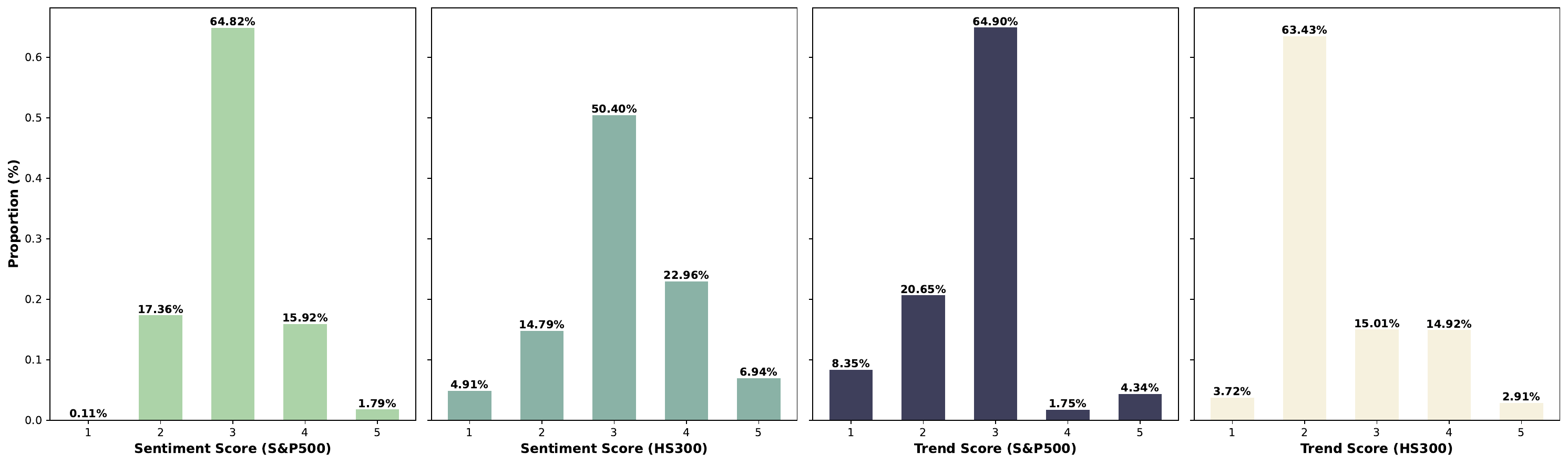}
        \vspace{-5mm} 
        \caption{Figures (a) and (b) show the proportions of LSA-generated news sentiment scores (1 = negative, 2 = somewhat negative, 3 = neutral, 4 = somewhat positive, 5 = positive) for S\&P 500 and HS300 stocks, respectively. Figures (c) and (d) display the corresponding six-month candlestick-chart trend scores using the same 1–5 scale (1 = negative trend, 5 = positive trend).}
        \label{fig:sentiment_trend_scores}
    \end{minipage}
    
\end{figure*}

For \textbf{the Chinese} market, daily \textbf{numerical} OHLCV data for HS 300 constituent stocks is retrieved through the Tushare API \footnote{\scriptsize\url{https://tushare.pro/}} and used to generate technical candlestick \textbf{charts} consistent with the U.S. market. \textbf{News} sentiment data is also collected from multiple Chinese financial media sources—including Sina Finance, Wallstreetcn, iFinD\footnote{\scriptsize
Sina Finance: \url{https://finance.sina.com.cn}; Wallstreetcn: \url{https://wallstreetcn.com}; iFinD: \url{https://www.ifind.com.cn}}; 
Eastmoney, YunCaijing\footnote{\scriptsize
Eastmoney: \url{https://www.eastmoney.com}; YunCaijing: \url{https://www.yuncaijing.com}}; 
Ifeng News, JRJ\footnote{\scriptsize
Ifeng News: \url{https://finance.ifeng.com}; JRJ: \url{https://www.jrj.com.cn}}
—covering the period from March 31, 2020, to March 31, 2025. A detailed overview of bilingual news is provided in Table \ref{Two News Sources}. Structured financial \textbf{table} data for the HS 300 is acquired via Tushare API, including quarterly and annual balance sheets, income statements, and cash flow statements for the period spanning 2005 to 2024.

\noindent\textbf{Data Ethics}
To ensure ethical compliance, we strictly adhere to robots.txt directives during news scraping, collecting only publicly accessible content that requires no payment or subscription.

\subsection{Data Preprocessing}
To construct FinMultiTime, we extract and align four distinct data modalities—technical charts, structured tables, normalized price series, and news —across mostly constituent stocks of the HS 300 and S\&P 500 indices, as of April 2025. The pipeline is designed to maximize temporal coverage while maintaining diversity in model inputs and ensuring comparability across the data sources.

\noindent\textbf{Technical Charts}  
For each stock, we segment daily OHLCV data—\\ Open, High, Low, Close, and Volume—into semi-annual windows and generate candlestick charts with corresponding volume bars. The original RGB charts are converted to 8-bit grayscale to reduce input dimensionality. We then prompt GPT-4.1 with a fixed instruction to assign one of five long-term trend categories to each chart: 1 (Slightly Up), 2 (Significantly Up), 3 (Flat), 4 (Slightly Down), or 5 (Significantly Down). This approach compresses multi-month price movement into a single trend label, serving as a visual indicator to enhance subsequent short-term price forecasting (Figure \ref{fig:image_prompt}).

\begin{table*}[t!]
\centering
\caption{Chinese (HS300) / English (S\&P500) Stock Time Series Data}
\vspace{-3mm} 
\renewcommand{\arraystretch}{1.1} 
\resizebox{0.85\textwidth}{!}{
\begin{tabular}{lccccccc}
\hline
\textbf{Date} & \textbf{Open} & \textbf{High} & \textbf{Low} & \textbf{Close} & \textbf{Volume} & \textbf{Dividends} & \textbf{Stock Splits} \\
\hline
2025-03-27 00:00:00 & 11.3700 & 11.4100 & 11.3500 & 11.3900 & 55334940 & 0 & 0 \\
2025-03-28 00:00:00 & 11.3900 & 11.4000 & 11.3400 & 11.3500 & 64494555 & 0.1275 & 1.2 \\
2025-03-31 00:00:00 & 11.3600 & 11.3800 & 11.2600 & 11.2600 & 111612564 & 0 & 0 \\
... & ... & ... & ... & ... & ... & ... & ... \\
\hline
\end{tabular}
}
\label{tab: ts}
\end{table*}

\begin{figure}[t!]
    \centering
    \begin{minipage}[b]{1\linewidth} 
        \centering
        \includegraphics[width=\linewidth]{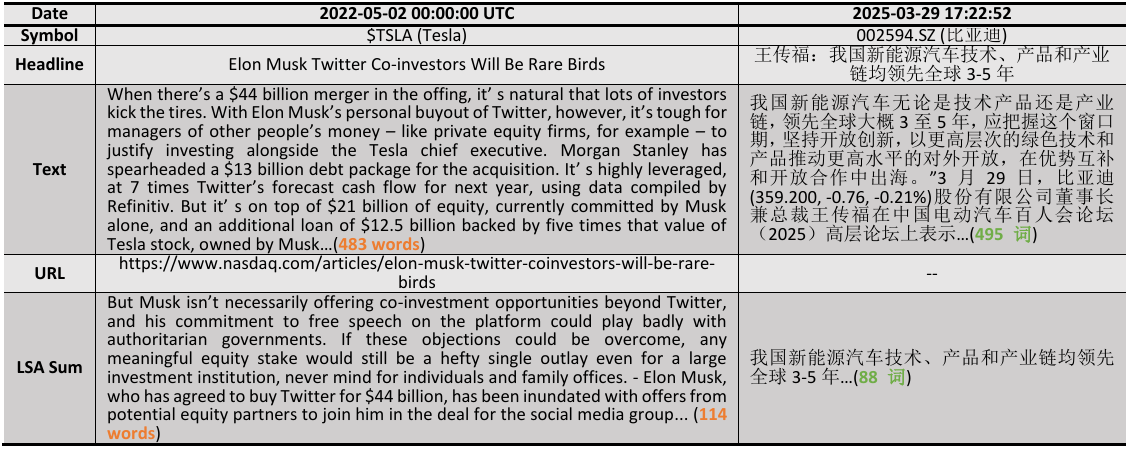}
        \vspace{-3mm}
        \caption{LSA-Generated Summaries of English and Chinese Stock News}
        \label{fig: lsa_summary}
    \end{minipage}
    \vskip -0.2in
\end{figure}

\begin{figure}[t!]
  \centering
  \begin{minipage}[b]{0.50\linewidth}
    \centering
    \includegraphics[width=\linewidth]{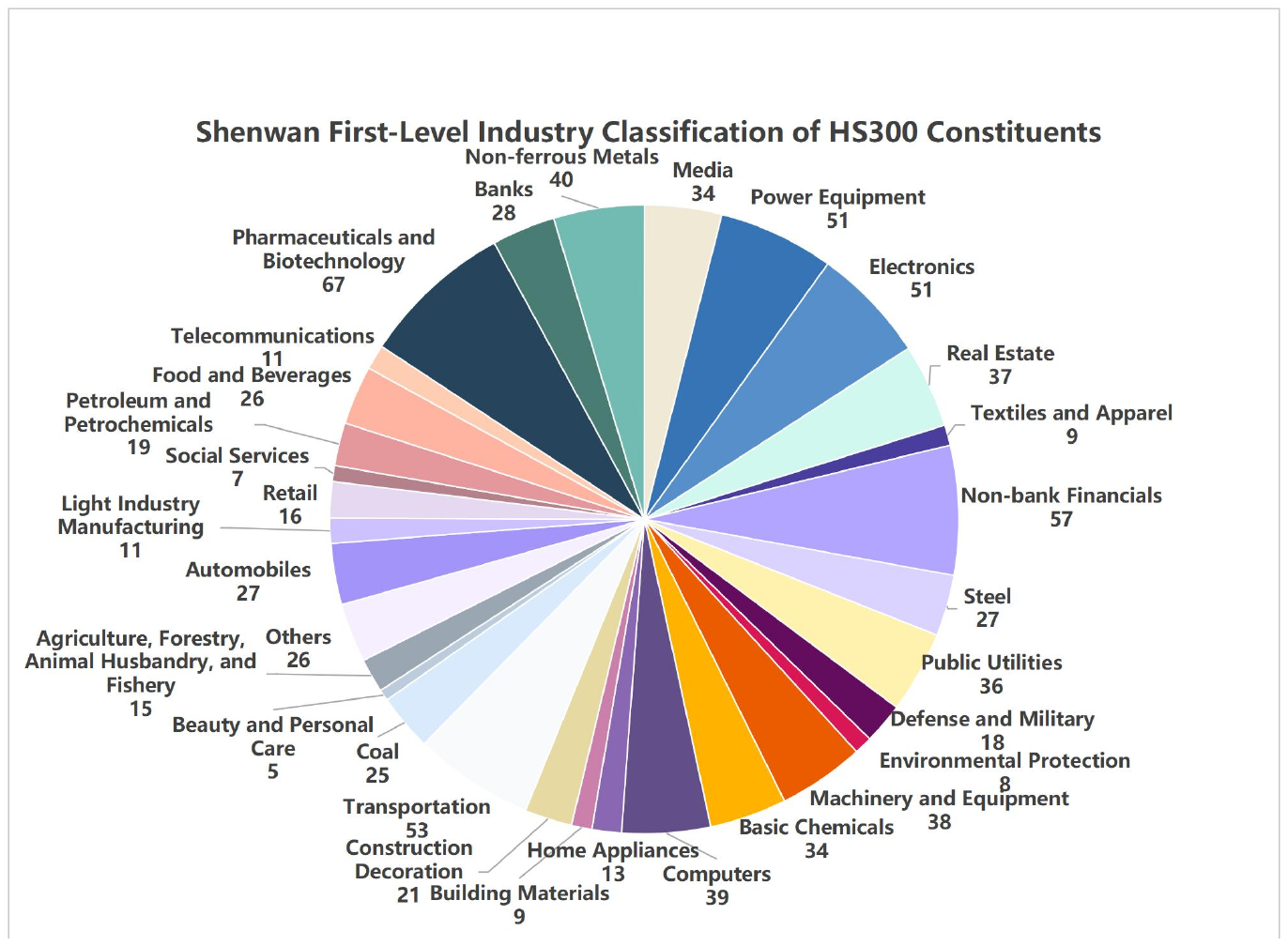}
    \caption{}   
    \label{fig:hs300pie}
  \end{minipage}
  \hfill
  \begin{minipage}[b]{0.45\linewidth}
    \centering
    \includegraphics[width=\linewidth]{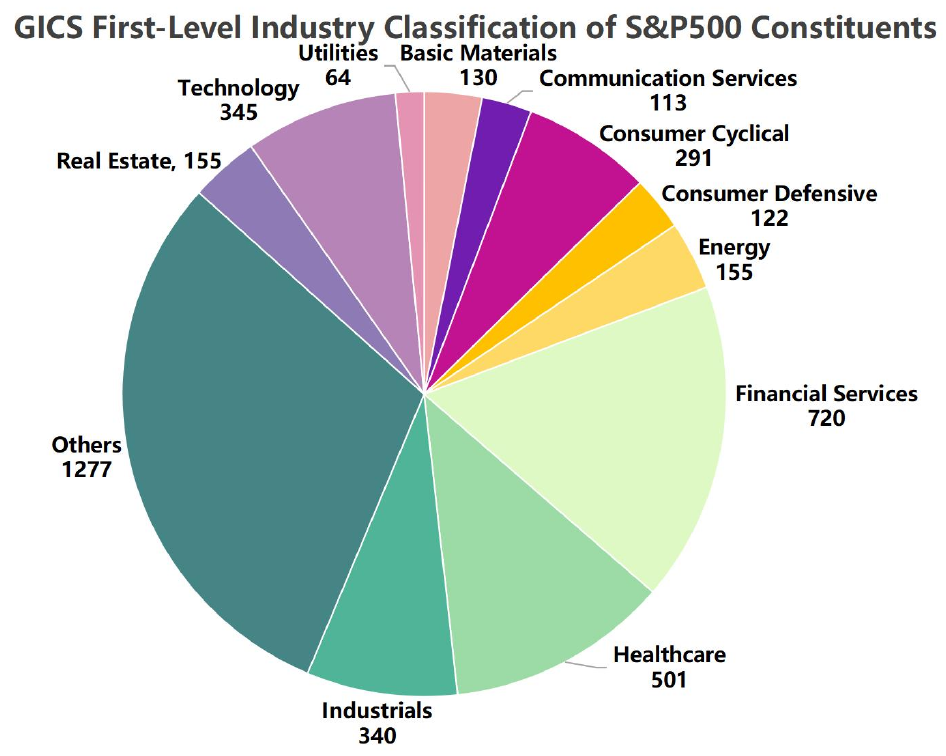}
    \caption{}  
    \label{fig:sp500pie}
  \end{minipage}
\end{figure}

\noindent\textbf{Structured Tables}  
For the structured-financial-table modality, we select six representative accounting variables that reflect profitability, liquidity, and capital structure. For HS300 stocks, we extract quarterly and annual values from the income statement and cash flow statement, including net profit, operating cash flow, and free cash flow. For S\&P 500 stocks, we collect analogous metrics from the balance sheet, cash flow statement, and statement of changes in equity, including shareholders’ equity, operating cash flow, and retained earnings (or accumulated deficit). All financial variables are aligned with the company’s reporting calendar. Period-end financial figures are matched to the closing price on the last trading day of each quarter or year, then forward-filled to cover all trading days within that reporting window, ensuring synchronization with daily price series.

\noindent\textbf{Price Series and News}  
Daily closing prices are normalized on a per-stock basis to enforce stationarity across both markets (Table \ref{tab: ts}). After collecting the raw URLs, headlines, and full texts, we use the \texttt{Sumy} library’s latent semantic analysis (LSA) algorithm to summarize each article into 3–4 sentences, which is approximately 16\% of the original length (Figure \ref{fig: lsa_summary}). A relevance weight \(W_{f}\) (see Appendix \ref{Bilingual News Summarize Algorithm}) is applied to prioritize sentences that mention the target stock ticker.  
To manage the volume of intraday news, we aggregate all summaries for a given stock on the same day, rank them by ticker frequency, and retain only the top entry as that day’s representative news. Each GPT-4.1 request includes at most ten entries (\emph{temperature}~=~0), ensuring deterministic sentiment inference. The model returns a sentiment score from 1 (negative) to 5 (positive). These scores are min-max normalized before multimodal fusion (Figure \ref{fig:news_prompt}).

\noindent\textbf{Summary}  
Figures \ref{fig:image_prompt} and \ref{fig:news_prompt} illustrate the five-level rubric for images and news, while Figure \ref{fig:sentiment_trend_scores} shows that the resulting sentiment distributions are approximately Gaussian. Mild skewness is observed: S\&P 500 scores are slightly left-skewed (mildly negative), while HS300 scores lean right (neutral to positive). This reflects broader market dynamics—U.S. softness versus a sustained rally in China during the sampling period.

\section{FinMultiTime Properties}

With data mining and preprocessing complete, FinMultiTime is now ready for in-depth analytical evaluation. This section highlights key insights drawn from a range of analytical perspectives.

\subsection{Dataset Overview}

FinMultiTime is a comprehensive and heterogeneous dataset exceeding 112.6 GB in total size. Its multidimensional structure highlights the dataset’s richness and diversity. The assembly process consumed approximately 5 TB of computing resources over a 60-day period, underscoring the complexity of the task and our commitment to ensuring high-quality, robust data for downstream analysis.

\subsection{Data Statistics}

\begin{figure*}[t!]
  \centering
    \includegraphics[width=\linewidth]{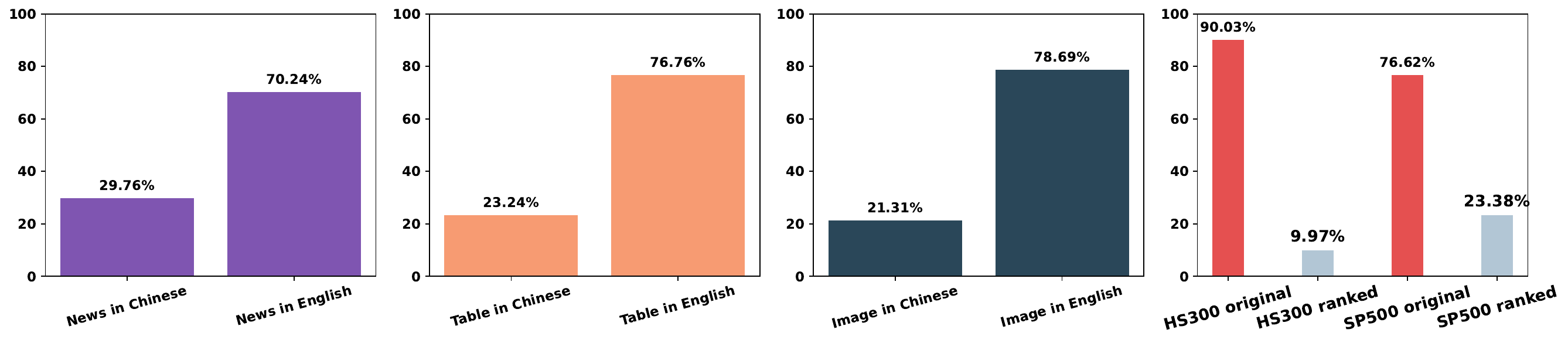}
    \vspace{-7mm}
    \caption{Proportions of Chinese vs. English Modalities (News, Tables, Images) and Coverage Ratios of Ranked vs. Original Daily News for HS300 and S\&P 500. }
    \label{fig:percentage}

\end{figure*}

\textbf{Language Distribution}
As shown in the first three panels of Figure~\ref{fig:percentage}, we compare the proportions of Chinese and English news articles, tabular records, and charts to illustrate FinMultiTime’s multilingual coverage and its relevance for global financial research.

\begin{figure*}[t!]
    \centering
    \begin{minipage}[b]{0.49\linewidth}
        \centering
        \includegraphics[width=\linewidth]{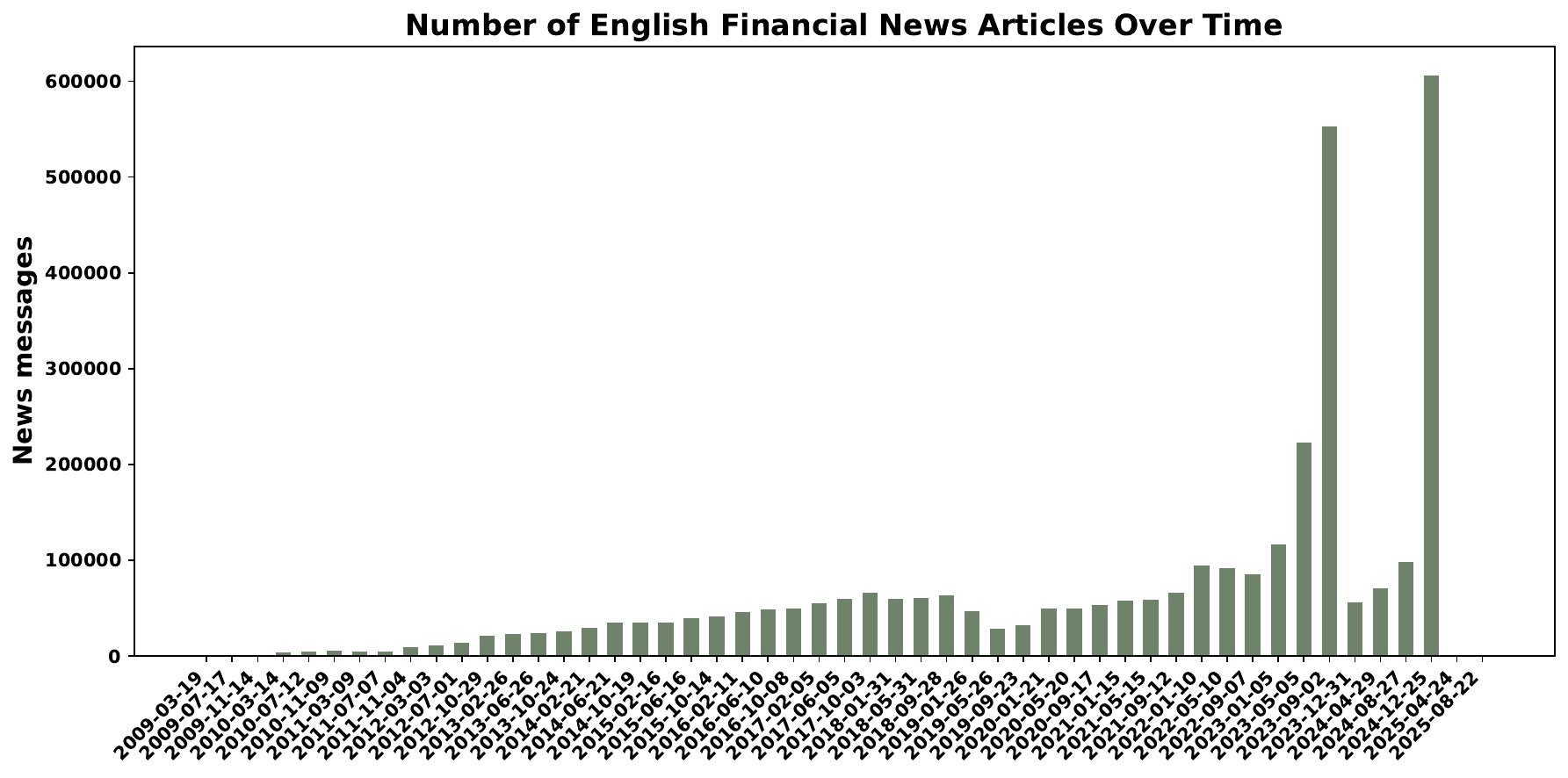}
        \vspace{-5mm}
        \caption{Number of S\&P500 News Articles Over Time}       
        \label{fig:sp500news_count}
    \end{minipage}
    \hfill
    \begin{minipage}[b]{0.49\linewidth}
        \centering
        \includegraphics[width=\linewidth]{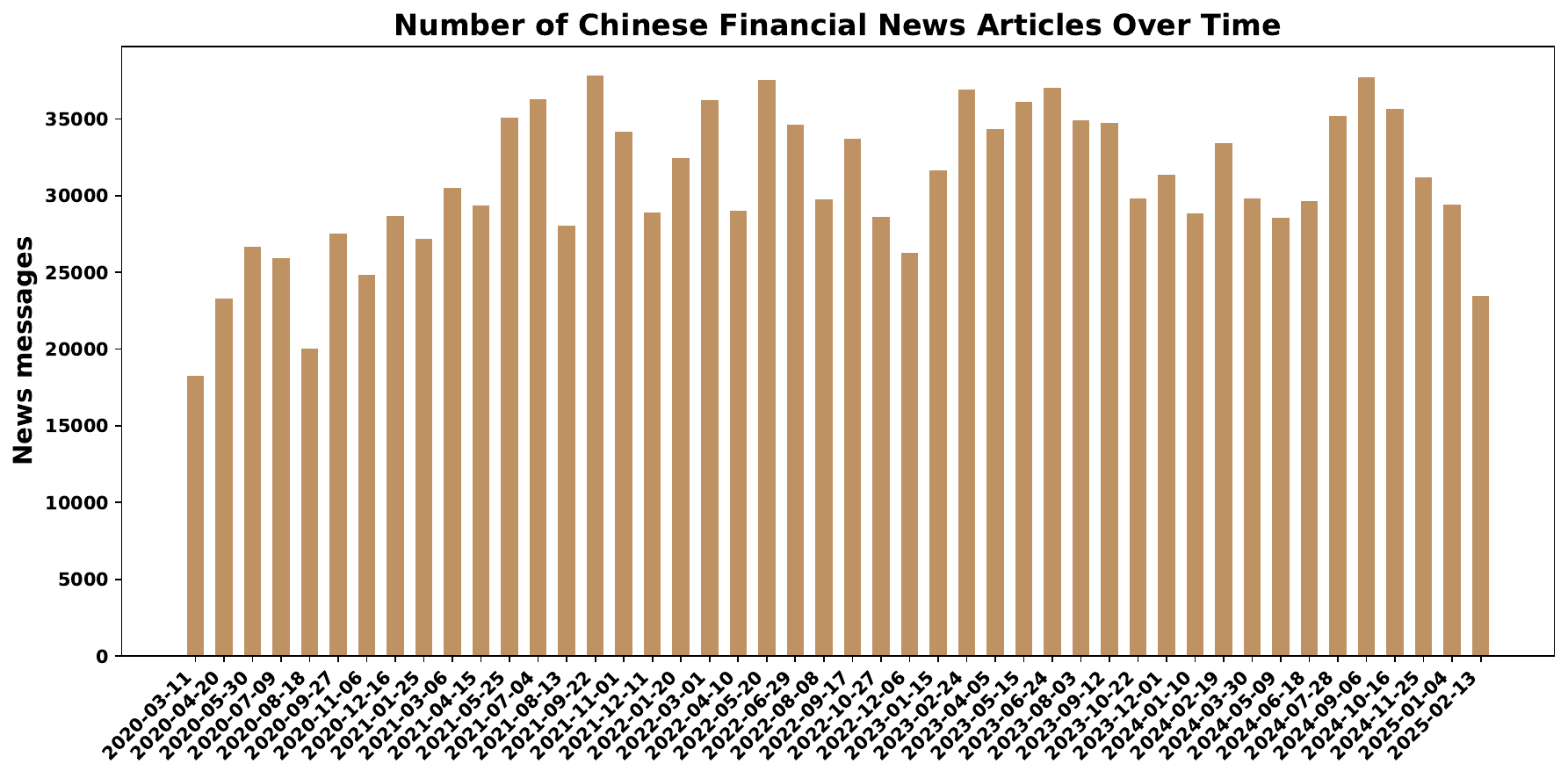}
        \vspace{-5mm}
        \caption{Number of HS300 News Articles Over Time}       
        \label{fig:hs300news_count}
    \end{minipage}
    \vspace{-3mm}
\end{figure*}

\begin{figure*}[t!]
    \centering
    \begin{minipage}[b]{0.49\linewidth}
        \centering
        \includegraphics[width=\linewidth]{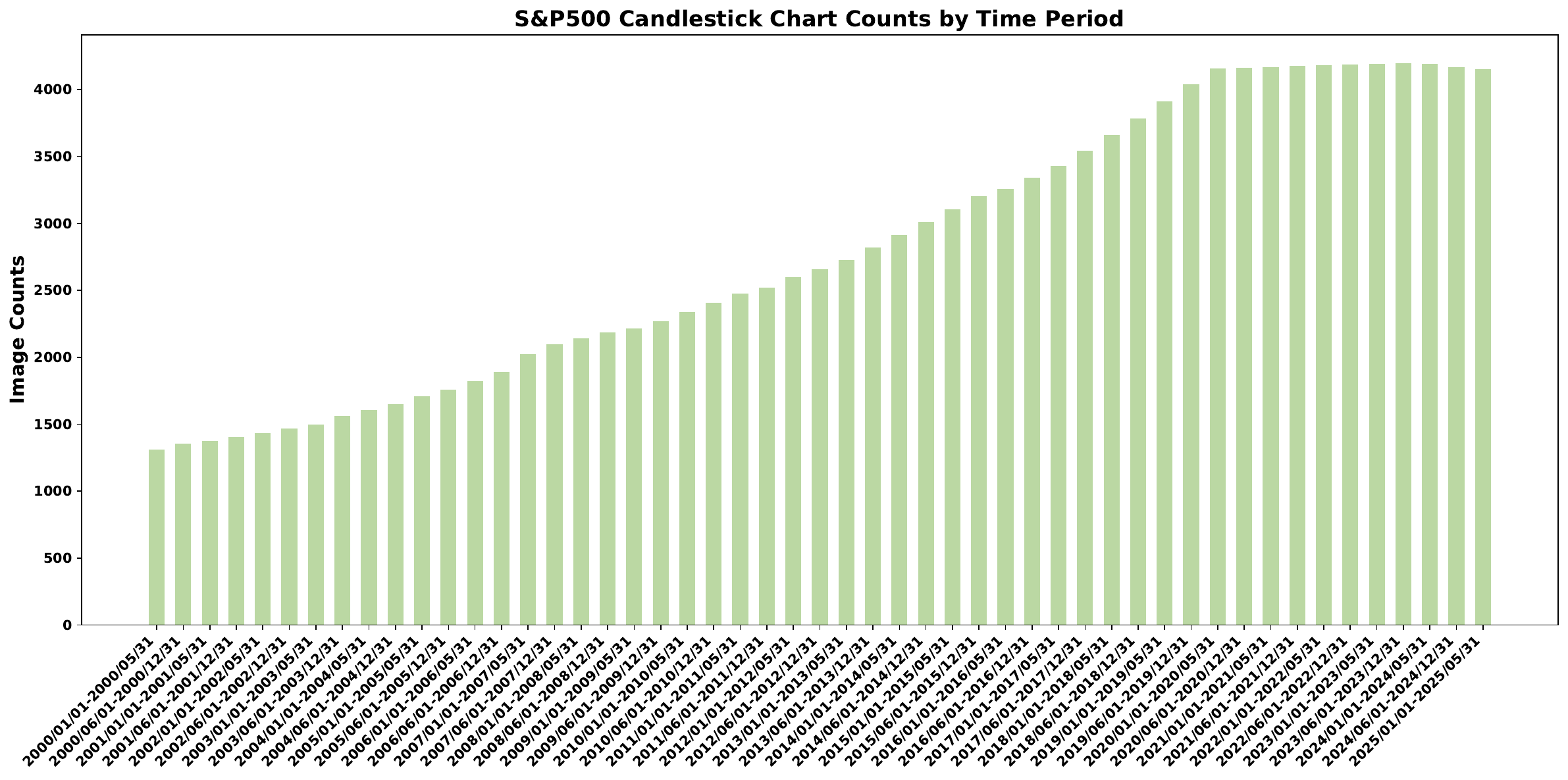}
        \vspace{-5mm}
        \caption{S\&P500 Candlestick Chart Counts by Time Period}       
        \label{fig:sp500_image_counts}
    \end{minipage}
    \hfill
    \begin{minipage}[b]{0.49\linewidth}
        \centering
        \includegraphics[width=\linewidth]{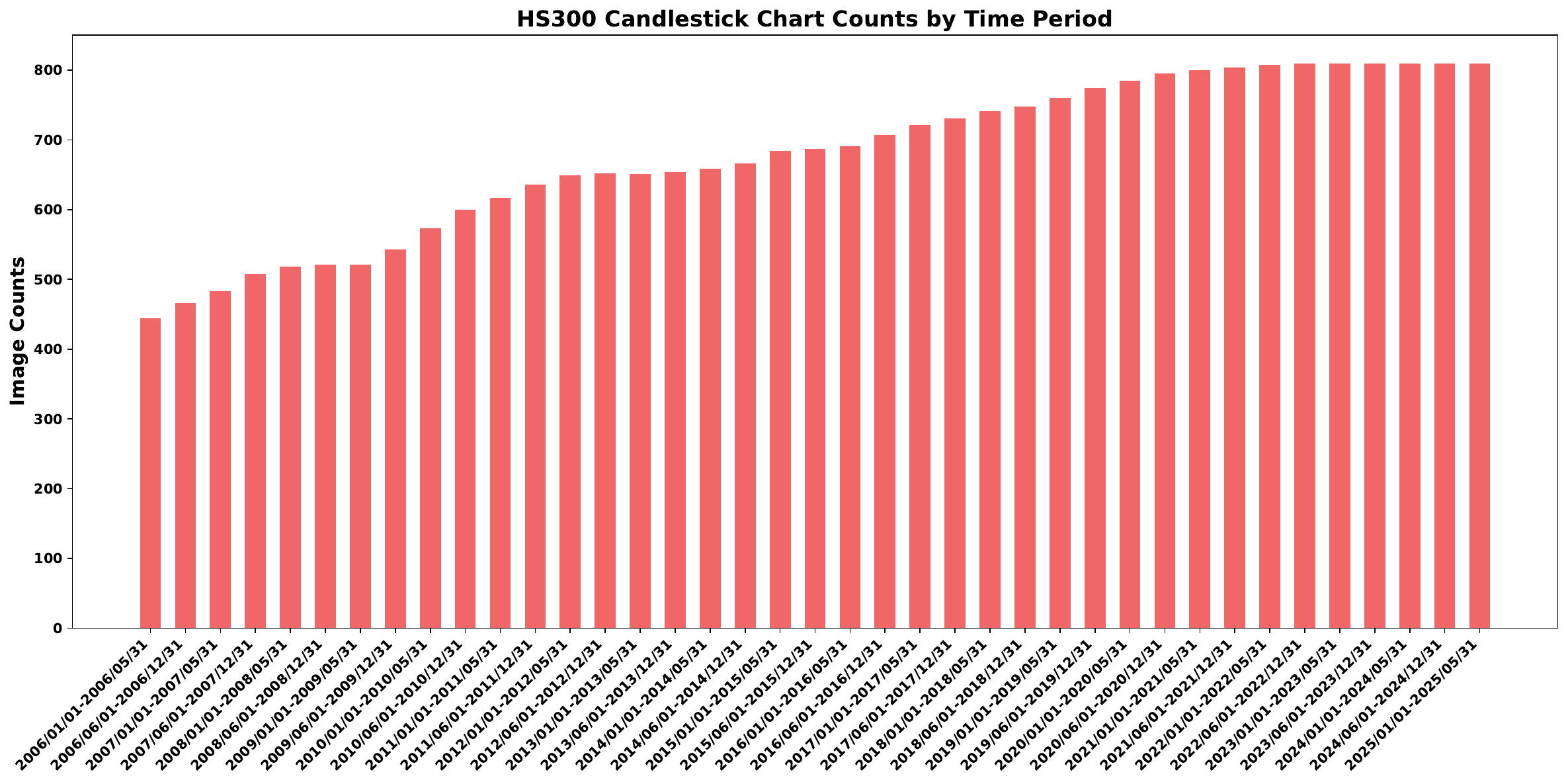}
        \vspace{-5mm}
        \caption{HS300 Candlestick Chart Counts by Time Period}       
        \label{fig:hs300_image_counts}
    \end{minipage}
\end{figure*}

\noindent\textbf{Temporal Distribution}
Figures~\ref{fig:sp500news_count} and~\ref{fig:hs300news_count} show the yearly volume of U.S. stock market news (1999–2025) and Chinese market news (2000–2025), respectively. Figures~\ref{fig:sp500_image_counts} and~\ref{fig:hs300_image_counts} present the number of K-line charts for the U.S. market (2006–2025) and the Chinese market (2000–2025). These temporal trends provide insights into the historical development of financial news coverage and visual analysis tools across markets.

\noindent\textbf{Industry Distribution}
Figures~\ref{fig:hs300pie} and~\ref{fig:sp500pie} compare the industry compositions of the HS300 constituents (classified by Shenwan Level-1) and the S\&P 500 constituents (classified by GICS Level-1). Under the more granular Shenwan classification, HS300 stocks are concentrated in sectors such as Pharmaceuticals \& Biotechnology, Non-Bank Financials, Transportation, Electrical Equipment, and Electronics. In contrast, the broader GICS classification highlights the dominance of large-scale sectors like Financial Services, Healthcare, Information Technology, and Industrials within the S\&P 500.

Together, these analyses illustrate FinMultiTime’s distinct value as a benchmark dataset for advanced financial analysis and time-series forecasting—thanks to its extensive market coverage, strong multilingual foundation, and broad temporal span.

\begin{table*}[t!]
\centering
\caption{HS300 vs.\ S\&P 500 — Multimodal Record Counts (35 stocks each)}
\vspace{-3mm}
\renewcommand{\arraystretch}{1.1}
\resizebox{\textwidth}{!}{%
\begin{tabular}{lcccc}
\hline
\textbf{} & \textbf{Semi‑annual trend images} & \textbf{Quarterly / annual tables} & \textbf{Daily time‑series points} & \textbf{News‑sentiment scores} \\
\hline
HS300    & 299,923 & 1,749 & 299,923 & 26,467 \\
S\&P 500 & 299,923 & 2,104 & 299,923 & 51,235 \\
\midrule
\textbf{Total} & 599,846 & 3,853 & 599,846 & 77,702 \\
\hline
\label{tab:35records}
\end{tabular}}%
    \vskip -0.2in
\end{table*}

\section{Experiments}

To validate the effectiveness of \emph{FinMultiTime}, we assessed the dataset’s overall performance through quantitative and qualitative evaluations. We outline our experimental strategy and demonstrate the dataset’s robustness in real‐world applications.

\begin{table*}[ht]
\centering
\caption{Model performance across modalities and prediction horizons (24, 48, 96) with fixed input length of 96 for \textcolor{blue}{HS300} stocks. Red values indicate the worst time-series prediction for each modality at the given horizon. The table compares five models over four modalities: (1) Time Series Only (price data), (2) News Sentiment, (3) Image Trend, and (4) Fundamental Table. Metrics reported are MAE and MSE (lower is better).}
\vspace{-3mm}
\renewcommand{\arraystretch}{1}
\resizebox{\textwidth}{!}{
\begin{tabular}{c|c|cc|cc|cc|cc|cc}
\hline
\multirow{3}{*}{\textbf{Horizon}} 
& \multirow{3}{*}{\textbf{Modality}}
& \multicolumn{2}{c|}{\textbf{CNN}} 
& \multicolumn{2}{c|}{\textbf{GRU}} 
& \multicolumn{2}{c|}{\textbf{LSTM}} 
& \multicolumn{2}{c|}{\textbf{RNN}} 
& \multicolumn{2}{c}{\textbf{TimeNet}} \\
\cline{3-12}
& & MAE (\(\downarrow\)) & MSE (\(\downarrow\)) 
  & MAE (\(\downarrow\)) & MSE (\(\downarrow\)) 
  & MAE (\(\downarrow\)) & MSE (\(\downarrow\)) 
  & MAE (\(\downarrow\)) & MSE (\(\downarrow\)) 
  & MAE (\(\downarrow\)) & MSE (\(\downarrow\)) \\
\hline
\multirow{4}{*}{\textbf{24}} 
& Time Series     & 0.1720 & \textcolor{red}{0.1097} & \textcolor{red}{0.2317} & \textcolor{red}{0.2523} & \textcolor{red}{0.0985} & \textcolor{red}{0.0281} & \textcolor{red}{0.2058} & \textcolor{red}{0.1197} & \textcolor{red}{0.0886} & \textcolor{red}{0.2808} \\
& News Sentiment  & 0.0756 & 0.0087 & 0.1101 & 0.0193 & 0.0834 & 0.0177 & 0.0707 & 0.0093 & 0.0268 & 0.1364 \\
& Fundamental Table & \textcolor{red}{0.1751} & 0.0582 & 0.0796 & 0.0117 & 0.0758 & 0.0104 & 0.1819 & 0.0501 & 0.0691 & 0.2425 \\
& Image Trend     & 0.0641 & 0.0082 & 0.0727 & 0.0104 & 0.0937 & 0.0202 & 0.0982 & 0.0201 & 0.0078 & 0.0721 \\
\hline
\multirow{4}{*}{\textbf{48}} 
& Time Series     & \textcolor{red}{0.2154} & \textcolor{red}{0.1219} & \textcolor{red}{0.4306} & \textcolor{red}{0.5236} & \textcolor{red}{0.1997} & \textcolor{red}{0.0721} & \textcolor{red}{0.3228} & \textcolor{red}{0.1831} & \textcolor{red}{0.1171} & \textcolor{red}{0.3064} \\
& News Sentiment  & 0.0770 & 0.0103 & 0.1955 & 0.0954 & 0.1130 & 0.0255 & 0.1149 & 0.0250 & 0.0736 & 0.2379 \\
& Fundamental Table & 0.1205 & 0.0269 & 0.1212 & 0.0225 & 0.1298 & 0.0235 & 0.1776 & 0.0446 & 0.0718 & 0.2508 \\
& Image Trend     & 0.0996 & 0.0165 & 0.1686 & 0.0531 & 0.1723 & 0.0570 & 0.1655 & 0.0352 & 0.0885 & 0.2701 \\
\hline
\multirow{4}{*}{\textbf{96}} 
& Time Series     & \textcolor{red}{0.2543} & \textcolor{red}{0.1481} & \textcolor{red}{0.6032} & \textcolor{red}{0.6982} & 0.1329 & 0.0311 & \textcolor{red}{0.1875} & \textcolor{red}{0.0633} & \textcolor{red}{0.1762} & \textcolor{red}{0.3917} \\
& News Sentiment  & 0.0899 & 0.0123 & 0.1841 & 0.0882 & 0.1257 & 0.0277 & 0.1689 & 0.0514 & 0.0901 & 0.2404 \\
& Fundamental Table & 0.1262 & 0.0397 & 0.1305 & 0.0261 & 0.1628 & 0.0430 & 0.1810 & 0.0457 & 0.1567 & 0.3375 \\
& Image Trend     & 0.1230 & 0.0399 & 0.1835 & 0.0537 & \textcolor{red}{0.1943} & \textcolor{red}{0.0881} & 0.1810 & 0.0626 & 0.0482 & 0.1751 \\
\hline
\end{tabular}
}
\label{tab:hs300}
\end{table*}

\begin{table*}[ht]
\centering
\caption{Model performance across modalities and prediction horizons (24, 48, 96) with fixed input length of 96 for \textcolor{blue}{S\&P500} stocks.}
\vspace{-3mm}
\renewcommand{\arraystretch}{1}
\resizebox{\textwidth}{!}{
\begin{tabular}{c|c|cc|cc|cc|cc|cc}
\hline
\multirow{3}{*}{\textbf{Horizon}} 
& \multirow{3}{*}{\textbf{Modality}} 
& \multicolumn{2}{c|}{\textbf{CNN}} 
& \multicolumn{2}{c|}{\textbf{GRU}} 
& \multicolumn{2}{c|}{\textbf{LSTM}} 
& \multicolumn{2}{c|}{\textbf{RNN}} 
& \multicolumn{2}{c}{\textbf{TimeNet}} \\
\cline{3-12}
& & MAE (\(\downarrow\)) & MSE (\(\downarrow\)) 
  & MAE (\(\downarrow\)) & MSE (\(\downarrow\)) 
  & MAE (\(\downarrow\)) & MSE (\(\downarrow\)) 
  & MAE (\(\downarrow\)) & MSE (\(\downarrow\)) 
  & MAE (\(\downarrow\)) & MSE (\(\downarrow\)) \\
\hline
\multirow{4}{*}{\textbf{24}} 
& Time Series     & \textcolor{red}{0.0498} & \textcolor{red}{0.0045} & \textcolor{red}{0.0949} & \textcolor{red}{0.0234} & \textcolor{red}{0.1120} & \textcolor{red}{0.0230} & \textcolor{red}{0.2158} & \textcolor{red}{0.1215} & \textcolor{red}{0.1085} & \textcolor{red}{0.3199} \\
& News Sentiment  & 0.0490 & 0.0039 & 0.0467 & 0.0041 & 0.0435 & 0.0036 & 0.0753 & 0.0102 & 0.0242 & 0.1329 \\
& Fundamental Table & 0.0463 & 0.0038 & 0.0931 & 0.0167 & 0.0888 & 0.0152 & 0.1664 & 0.0395 & 0.0894 & 0.2910 \\
& Image Trend     & 0.0475 & 0.0041 & 0.0525 & 0.0055 & 0.0723 & 0.0122 & 0.1485 & 0.0301 & 0.0052 & 0.0594 \\
\hline
\multirow{4}{*}{\textbf{48}} 
& Time Series     & \textcolor{red}{0.0820} & \textcolor{red}{0.0151} & \textcolor{red}{0.1433} & \textcolor{red}{0.0445} & \textcolor{red}{0.0895} & \textcolor{red}{0.0166} & \textcolor{red}{0.1754} & \textcolor{red}{0.1104} & \textcolor{red}{0.1646} & \textcolor{red}{0.3913} \\
& News Sentiment  & 0.0559 & 0.0053 & 0.0569 & 0.0059 & 0.0880 & 0.0159 & 0.1148 & 0.0218 & 0.1089 & 0.3147 \\
& Fundamental Table & 0.0515 & 0.0051 & 0.1207 & 0.0280 & 0.0868 & 0.0125 & 0.0806 & 0.0099 & 0.1320 & 0.3587 \\
& Image Trend     & 0.0714 & 0.0091 & 0.0708 & 0.0118 & 0.0799 & 0.0165 & 0.1452 & 0.0294 & 0.0970 & 0.2998 \\
\hline
\multirow{4}{*}{\textbf{96}} 
& Time Series     & 0.0769 & 0.0095 & \textcolor{red}{0.2251} & \textcolor{red}{0.1129} & 0.1380 & 0.0322 & \textcolor{red}{0.1183} & 0.0209 & \textcolor{red}{0.2027} & \textcolor{red}{0.4381} \\
& News Sentiment  & 0.0592 & 0.0055 & 0.0937 & 0.0178 & 0.1259 & 0.0266 & 0.1129 & \textcolor{red}{0.0213} & 0.0895 & 0.2643 \\
& Fundamental Table & 0.0775 & 0.0105 & 0.1163 & 0.0228 & \textcolor{red}{0.1465} & \textcolor{red}{0.0356} & 0.1158 & 0.0182 & 0.1846 & 0.4098 \\
& Image Trend     & \textcolor{red}{0.0939} & \textcolor{red}{0.0201} & 0.1174 & 0.0340 & 0.1110 & 0.0312 & 0.1069 & 0.0149 & 0.0482 & 0.1937 \\
\hline
\end{tabular}
}
\label{tab:sp500}
\end{table*}

\begin{table}[ht]
\centering
\caption{
Performance comparison of three state-of-the-art bimodal models—Chat Time, CALF, and FTS-Text-MoE—using both text and time-series inputs. Evaluations are conducted on two datasets (S\&P 500 and HS 300) across three stock subsets (5, 15, and 35 stocks). Prediction horizon is 24 with a fixed input length of 96. Best scores per row are highlighted.
}
\vspace{-3mm}
\resizebox{\linewidth}{!}{
\begin{tabular}{c|c|cc|cc}
\hline
\multirow{2}{*}{\textbf{\#}} & \multirow{2}{*}{\textbf{Model}} 
& \multicolumn{2}{c|}{\textbf{S\&P 500}} 
& \multicolumn{2}{c}{\textbf{HS 300}} \\
\cline{3-6}
& & MAE (\(\downarrow\)) & MSE (\(\downarrow\)) & MAE (\(\downarrow\)) & MSE (\(\downarrow\)) \\
\hline
\multirow{3}{*}{\textbf{5}} 
& Chat Time     & 0.4753 & 0.4432 & 0.2647 & \textbf{0.1195} \\
& CALF          & 0.3763 & 0.2688 & 0.5707 & 0.6290 \\
& FTS-Text-MoE & \textbf{0.0894} & \textbf{0.2263} & \textbf{0.2506} & 0.3789 \\
\hline
\multirow{3}{*}{\textbf{15}} 
& Chat Time     & 0.6648 & 0.9103 & 0.4804 & \textbf{0.3975} \\
& CALF          & 0.4996 & 0.5649 & 0.4921 & 0.5660 \\
& FTS-Text-MoE & \textbf{0.3299} & \textbf{0.4818} & \textbf{0.3232} & 0.4651 \\
\hline
\multirow{3}{*}{\textbf{35}} 
& Chat Time     & 0.6431 & 0.6928 & 0.5643 & 0.5117 \\
& CALF          & 0.5503 & 0.6011 & 0.5019 & 0.5778 \\
& FTS-Text-MoE & \textbf{0.3914} & \textbf{0.5344} & \textbf{0.3817} & \textbf{0.5010} \\
\hline
\end{tabular}
}
\vspace{-0.3cm}
\label{tab:sp500_hs300}
\end{table}

\subsection{Experiment Settings}
We evaluate stock prediction performance on a FinMultiTime subset of 70 representative stocks, comparing various approaches.
\subsubsection{Datasets}
We conduct our experiments using a subset of the FinMultiTime dataset, specifically selecting the 35 most influential constituents of the 2025 S\&P 500 index and the 35 most influential constituents of the HS300 index, resulting in a total of 70 representative stocks. These samples were processed through our annotation pipeline, producing 77,702 sentiment-annotated news articles, 599,846 semiannual K-line charts, and 3,853 quarterly or annual structured financial records. For detailed statistics, please refer to Table \ref{tab:35records}.

\subsubsection{Compared Methods}
We conducted a comparative study of existing unimodal and bimodal approaches for stock price prediction. 

\noindent\textbf{Traditional time-series models:} These methods transform each modality—historical prices, news articles, financial tables, and K-line charts—into temporal sequences. This includes price histories, sentiment scores derived from news, financial variables extracted from tables, and visual features obtained from charts. Each resulting sequence is then processed individually by standard architectures such as RNN, LSTM, GRU \citep{shen2018deep}, 1D CNN \citep{chen2016financial}, and a 4-layer TimeNet \citep{wu2022timesnet}.

\noindent\textbf{Recent bimodal models combining text and time series:} We also evaluate three representative LLM-based approaches that jointly utilize text and time-series data: CALF \citep{liu2025calf}, ChatTime \citep{wang2025chattime}, and FTS-Text-MoE \citep{xu2025learning}. CALF enhances cross-modal fusion by aligning data distributions; ChatTime treats time series as language tokens to enable unified LLM modeling; and FTS-Text-MoE leverages a sparse, lightweight MoE Transformer decoder for efficient forecasting. All three methods take raw news text and structured time-series data as joint inputs.

\subsection{Qualitative Tests}

\noindent\textbf{Impact of Multimodal Data on Prediction Performance:} Tables \ref{tab:hs300} and \ref{tab:sp500} demonstrate clearly that incorporating additional modalities such as news sentiment, fundamental data, and image trends substantially enhances model prediction performance. For instance, in predicting HS300 stocks with a 24-hour horizon using the GRU model, the mean squared error (MSE) drops significantly from 0.2523 (time-series only) to 0.0727 upon including the image trend modality. Similarly, for S\&P500 stocks with a 48-hour horizon using the CNN model, the MSE improves notably from 0.0151 with pure time-series data to 0.0053 when news sentiment data is added. These examples illustrate that leveraging multimodal data effectively enhances prediction accuracy for stock prices.

\noindent\textbf{Comparative Analysis of Model Performance:} A comparison of the two tables highlights distinct differences in performance among models across varying prediction horizons. For example, in predicting HS300 stocks with a 96-hour horizon, the TimeNet model achieves a mean absolute error (MAE) of 0.1762 using only time-series data, whereas the CNN model performs less effectively, showing an MAE of 0.2543. This indicates TimeNet's superior capability for longer-term forecasting. Conversely, in the 24-hour horizon prediction task for S\&P500 stocks, the CNN model exhibits an MSE of 0.0045 using only time-series data, significantly outperforming both the RNN model (MSE of 0.1215) and the TimeNet model (MSE of 0.3199). This clearly indicates the CNN model’s superior performance in short-term forecasting scenarios. Therefore, selecting an appropriate model based on the forecast horizon and data modality is critical for achieving optimal prediction performance.

\subsection{Quantitative Tests}
To investigate the impact of dataset scale on performance, we trained models using bilingual datasets (HS300/S\&P 500), each comprising three subsets with 5, 15, and 35 stocks.  

As shown in Table \ref{tab:sp500_hs300}, model performance consistently improves as the number of stocks increases from 5 to 35, underscoring the benefits of larger training datasets for enhancing predictive accuracy. Furthermore, among the three evaluated models (Chat Time, CALF, and FTS-Text-MoE), the \textsc{FTS-Text-MoE} model achieves the best performance in most cases. Specifically, for the S\&P 500 dataset with 5 stocks, \textsc{FTS-Text-MoE} attains a significantly lower mean absolute error (MAE) of 0.0894 and mean squared error (MSE) of 0.2263, compared to Chat Time’s MAE of 0.4753 and MSE of 0.4432, and CALF’s MAE of 0.3763 and MSE of 0.2688. Similarly, in the HS300 dataset with 35 stocks, the \textsc{FTS-Text-MoE} model records the lowest MAE (0.3817) and MSE (0.5010), clearly surpassing the performance of Chat Time (MAE 0.5643, MSE 0.5117) and CALF (MAE 0.5301, MSE 0.5718).  

These results highlight the superior capability of the \textsc{FTS-Text-MoE} model in effectively integrating textual and time-series data to capture complex patterns, thereby yielding more accurate short-term stock price predictions. All experiments utilized 96 days of historical data to forecast prices for the subsequent 24 days. Each model was trained for 100 epochs, evaluated on the 5-stock split after removing one outlier, and the reported results represent mean values.

\section{Related Work}
\noindent\textbf{Financial time-series models} Traditional time-series models like linear regression \cite{weisberg2005applied}, ARIMA \cite{ariyo2014stock} and GARCH \cite{bauwens2006multivariate} depend on stationarity and strong assumptions, so they often miss complex dependencies or abrupt shocks. Recently, machine learning \cite{krollner2010financial, kelly2023financial, kim2003financial}, deep learning \cite{sezer2020financial} and NLP \cite{xing2018natural, lukauskas2022economic} have tapped sentiment and other qualitative signals to enhance forecast accuracy. This trend mirrors Markowitz’s market correlation concept, linking sentiment from news, blogs, and social media to asset prices. With growing data and compute, LLMs now enable finer sentiment quantification \cite{lopez2023can}. Moreover, TSMixer-MICM \cite{koval2024financial} turns quarterly financial-statement tables into time-series features, aligning them with price and text data for three-modal analysis.

\noindent\textbf{Financial Multimodal time‑series datasets.} Financial Multimodal time‑series datasets fall into two groups. General economic collections (e.g., Time‑MMD \cite{liu2024time}, CiK \cite{williams2024context}) pair macro‑text with monthly indicators but are too coarse and small for fine‑grained forecasting. Financial‑specific sets target asset prices: NewsForecast links Bitcoin news to daily prices; TimeCAP\cite{lee2025timecap}, DOW30\cite{chen2023chatgpt}, TSQA\cite{kong2025time} align stock news with prices; ACL18\cite{xu2018stock}, CIKM18\cite{wu2018hybrid}, SEP\cite{koa2024learning} use tweet sentiment. FinBEN \cite{xie2024finben} and FNSPID Nasdaq \cite{dong2024fnspid} add bilingual text yet remain text‑price only, while a 2024 EMNLP Findings study\cite{koval2024financial} is first to fuse quarterly tables with text and prices, albeit at low frequency and small scale. Overall, these datasets are modest in size and mostly single‑market (chiefly U.S.), limiting their usefulness for pre‑training and evaluating emerging large‑scale financial LLMs and multimodal models. 

\section{Conclusion}
In this work, we present FinMultiTime, the first large-scale, cross-market multimodal dataset for financial time-series forecasting. By temporally aligning stock prices, financial news, structured fundamentals, and K-line charts across the S\&P 500 and HS300 universes, FinMultiTime provides a rich foundation for AI-driven financial prediction. Our experiments highlight three key findings. First, larger dataset scale consistently enhances model accuracy, confirming the importance of data volume for robust learning. Second, multimodal fusion significantly boosts predictive performance, as evidenced by substantial error reductions when incorporating sentiment, fundamentals, or image-based trends. Third, comparative analysis reveals that model choice and prediction horizon strongly influence outcomes. Finally, the reproducible pipeline ensures ongoing dataset expansion and adaptability, enabling the research community to build upon this resource. Overall, FinMultiTime advances financial forecasting by integrating heterogeneous signals at scale.

\clearpage

\bibliographystyle{ACM-Reference-Format}
\bibliography{sample-base}

\appendix

\appendix

\section{Bilingual News Summarize Algorithm}
\label{Bilingual News Summarize Algorithm}
In reference to FNSPID \cite{dong2024fnspid}, we introduce a weight model \( W_z \) to enhance summarization and emphasize relevant stocks. In the \texttt{sumy} package, all terms are included in the summary. Exclusiveness involves rephrasing sentences rather than extracting terms. We parse the graph \( G \) into sentences and assign a weight \( W_p \) based on relevance to the stock symbol, setting \( k = 1 \) for sentences containing the symbol. For summarized sentences \( S_{sum} \), a score of \( t = 1 \) is given if the sentence is longer. In Equation (6), we combine \( W_p \) and \( W_q \) to calculate the final weight \( W_z \), with irrelevant sentences receiving a weight of 0. The sentences are sorted by weight to form the final summary.

\[
W_p(S, s) = \begin{cases} 
k & \text{if } S \in G \\
0 & \text{otherwise}
\end{cases}
\]

\[
W_q(S_{sum}, S_{long}) = \begin{cases} 
t & \text{if } S_{sum} \in S_{long} \\
0 & \text{otherwise}
\end{cases}
\]

\[
W_z = W_p + W_q
\]


\section{FinMultiTime Applications} 
This section critically examines the potential uses of the FinMultiTime dataset in financial‑market research, the technical hurdles encountered during its construction, and the attendant ethical challenges, while outlining avenues for future work.

\subsection{Construction Challenges}
Bilingual news extraction and sentiment labelling. We experimented with lightweight extractive algorithms (Luhn, LexRank, TextRank) and generative models (distilbart‑cnn‑12‑6). Although both approaches handle simple sentiments (e.g., “sharp price rise” or “steep decline”) reasonably well, extractive methods often miss key context in longer passages, whereas generative models suffer from summary repetition, unstable scores, and attention drift on lengthy documents.

Modal imbalance. Relying on a small set of tabular variables or on trend labels derived solely from candlestick images fails to unlock the complementary value of FinMultiTime’s four modalities. These limitations underscore the need for more efficient architectures that can exploit mutual information among text, images, and structured data to reveal genuine predictive power.

\subsection{Prospective Use Cases}
\textbf{Multimodal model training.} The temporally aligned text, numeric, image, and table streams enable the development of joint‑learning models for stock prediction. Such models can bolster robustness to short‑horizon noise and improve reinforcement‑learning agents in sequential decision‑making—especially for trend forecasting and strategy design.

\textbf{Sentiment and trend‑signal analysis.} Combining news‑sentiment scores with long‑horizon trend labels allows researchers to assess the incremental explanatory power of non‑price signals within a modern portfolio‑theory framework. Batch processing of sentiment and trend indicators across many tickers further refines market forecasts and portfolio allocation.

\textbf{Correlation and anomaly detection.} The four aligned modalities facilitate granular studies of how sentiment, image‑based trends, and fundamentals correlate with price dynamics, potentially revealing latent market drivers. Pattern matching on historical data can surface precursors of systemic risk, offering fresh tools for volatility warnings and risk management.

\textbf{Financial generative‑AI applications.} With its large, heterogeneous corpus, FinMultiTime serves as prime fine‑tuning material for large language models, powering next‑generation robo‑advisers, automated report writers, and other finance‑oriented AI services.

\subsection{Ethical Considerations}
Privacy and data security. Financial records often contain sensitive personal or institutional information. We employ state‑of‑the‑art anonymisation and de‑identification techniques and adhere strictly to GDPR, CCPA, and related regulations to safeguard privacy throughout data collection and processing.

\textbf{Misuse risks.} Predictive models built on FinMultiTime could be misappropriated, leading to market manipulation or systemic risk. We therefore conduct bias and fairness audits and publish explicit usage guidelines to curb discriminatory or misleading outcomes.

\textbf{Transparency and traceability.} Every record is source‑tagged, and detailed processing documentation is released publicly, ensuring reproducibility, auditability, and responsible research practice.

By addressing construction bottlenecks, enriching multimodal use cases, and enforcing rigorous ethical safeguards, FinMultiTime not only provides a solid empirical foundation for financial‑market analysis but also sets a high academic and ethical benchmark for future industry and scholarly endeavours.

\section{Future Work}

Expanding the FinMultiTime Dataset: Although our coverage of stock‑related data is extensive, financial data remain inherently time‑sensitive. We plan to develop an automated pipeline to continuously ingest and update news feeds, thereby substantially enlarging the dataset’s scope and currency.

\textbf{Unlocking FinMultiTime’s Full Potential:} As the most comprehensive resource aligning price series, sentiment annotations, long-term trend signals, and corporate fundamental data, FinMultiTime can support several frontier research directions:

\textbf{Multimodal Modeling:} Multimodal modeling will integrate heterogeneous sources—text, images, tables, and time series—to construct more robust market‐prediction models; sentiment‐impact analysis will quantitatively assess how news sentiment drives stock‐price volatility, thereby advancing sentiment‐analysis algorithms; trend‐signal evaluation will investigate the contribution of long‐term trend indicators to forecasting accuracy; and fundamental‐data integration will examine the auxiliary role of financial‐statement features in investment decision‐making to enhance real‐world applicability. Although our news coverage is already extensive, the synergistic exploitation of chart images, textual summaries, and tabular data remains underexploited. In future work, we will explore pre‐training language models within a reinforcement‐learning framework to improve multimodal feature extraction and its downstream applications.

By identifying these limitations and outlining targeted research avenues, we aim to inspire subsequent studies and further enhance the value and impact of the FinMultiTime dataset.

\end{document}